\documentclass[aps,prc,reprint,showpacs,floatfix,superscriptaddress]{revtex4-2}
\usepackage{graphicx} % include figure files
\usepackage{dcolumn} % align table columns on decimal point
\usepackage{bm} % bold math
\usepackage{color}

\usepackage{float}

\begin{document}

\title{ Schiff moments of deformed nuclei.
}

\author{O. P. Sushkov}
%\email{}
\affiliation{
  School of Physics, The University of New South Wales, Sydney, New South Wales
  2052, Australia
}

% \pacs{75.30.Ds, 75.40.Gb, 75.50.Ee, 05.10.Cc}

\date{\today}

\begin{abstract}
  Stimulated by recent suggestion of Cosmic Axion Spin Precession Experiment
  with Eu contained compound we develop a new method for accurate calculation
  of Schiff moments of even-odd deformed nuclei.
  The method is essentially based on experimental data  on magnetic moments and
  E1,E3-amplitudes in the given even-odd nucleus and in adjacent even-even
  nuclei. Unfortunately such sets of data are not known yet for most of
  interesting
  nuclei. Fortunately the full set of data is available for $^{153}$Eu.
  Hence, we perform the calculation for $^{153}$Eu and find value of the
  Schiff moment.
  The value is about 30 times larger than a typical  Schiff moment
  of a spherical heavy nucleus. The  enhancement of the Schiff moment
  in   $^{153}$Eu  is related to  the low energy octupole mode.
  On the other hand the value of Schiff moment  we find is 30 times
  smaller than that  obtained in the assumption of static octupole
  deformation.
  \end{abstract}
\maketitle

\section{Introduction}
Electric dipole moment (EDM) of an isolated quantum object in a nondegenerate
quantum state is a manifestation of violation of time reversal (T) and
parity (P) fundamental symmetries. Searches of EDM
of neutron is a long quest for fundamental
P,T-violation~\cite{Ramsey1982,Serebrov2015,Abel2020}.
EDM of a nucleus can be significantly larger than that of a
neutron~\cite{Sushkov1984}.
However a nucleus has nonzero electric charge and therefore in a charge neutral
system (atom, molecule, solid) EDM of nucleus cannot be measured~\cite{Schiff1963}. The quantity that can be measured is the so called Schiff Moment (SM)
which is nonzero due to the finite nuclear size~\cite{Sushkov1984}.
Like EDM the SM is a vector directed along the angular momentum.

Renewal of my interest to this problem is related to Cosmic Axion Spin
Precession Experiment (CASPEr) on searches of the QCD axion dark matter.
The current CASPEr experiment is based on Lead Titanate
ferroelectric~\cite{Budker2014}, see also Ref.~\cite{Mukhamedjanov2005}.
The experiment is probing the Schiff moment of $^{207}Pb$ nucleus.
There is a recent suggestion~\cite{ASushkov2023a} to use for CASPEr
experiment the crystal of EuCl$_3\cdot$ 6H$_2$O instead of Lead Titanate.
The major advantage is experimental: a possibility to polarise Eu nuclei via
optical pumping in this crystal
allows to improve sensitivity by orders of magnitude.
Expected effect in  EuCl$_3\cdot$ 6H$_2$O has been calculated in
Ref.~\cite{ASushkov2023}.  
The observable effect in a solid is built like a Russian doll Matreshka,
it has four different spatial and energy scales inside each other.
(i) Quark-gluon scale, $r < 1$fm,
(ii) Nuclear scale, $1fm \lesssim r \lesssim  10fm $,
(iii) Atomic scale, $10fm <  r \lesssim  1\AA $,
(iv) Solid state scale, $r > 1\AA$.
The calculation~\cite{ASushkov2023} is pretty accurate at the scale (iii),
it has an uncertainty at most by factor 2 at the scales (i) and (iv).
However, the uncertainty at the scale (ii), the nuclear scale, is two
orders of magnitude, this is the uncertainty in $^{153}$Eu Schiff moment.
Such an uncertainty is more or less typical for deformed even-odd nuclei.
The aim of the present work is twofold (i) development of the accurate method
for SM calculation, (ii) performamce of the calculation for $^{153}$Eu.
A reliable purely theoretical calculation is hardly possible.
Therefore, our appoach is to use available experimental
data as much as posible.

$^{153}$Eu is a deformed nucleus. A simplistic estimate of SM of a
nucleus with quadrupolar deformation based on Nilsson model
performed in Ref.~\cite{Sushkov1984}
gave a result by an order of magnitude larger than SM of a spherical heavy
nucleus, say SM of $^{207}Pb$.
It has been found later in Ref.~\cite{Auerbach1996} that if the nucleus
has a static octupolar deformation the SM is dramatically enhanced.
Based on analysis of rotational spectra of $^{153}$Eu  authors of
Ref.~\cite{Flambaum2020} argued that $^{153}$Eu has a static octupolar
deformation and hence, using the idea~\cite{Auerbach1996}
arrived to the estimate of SM that is $10^3$ times larger than that of a
heavy spherical nucleus.

To elucidate structure of wave functions of $^{153}$Eu in the 
present work we analyse available experimental data on magnetic moments and
amplitudes of E1,E3-transitions. In the result of this
analysis we confidently claim that the model of static octupolar deformation
is too simplistic. Nilsson wave functions of quadrupolar deformed nucleus are
almost correct. However, this does not imply that the octupolar mode is
irrelevant.
There is an admixture of the octupole vibration to the Nilsson states
and we determine the amplitude of the admixture. All in all this allows us
to perform a pretty reliable and accurate calculation of SM.

To avoid misundertanding, our statement about the magnitude of the SM
is based on analysis of a broad set of data, therefore, the statement is
nuclear
specific, it is valid for $^{153}$Eu and it is valid for $^{237}$Np.
Unfortunately such sets of data are not known yet for many interesting
nuclei.

Structure of the paper is the following.
In Section II we analyse lifetimes of relevant levels in $^{152}$Sm and
$^{153}Eu$ and hence find the relevant E1-amplitudes.
The Section III is the central one, here we discuss the structure of wave
functions of the parity doublet $|5/2^{\pm}\rangle$ in $^{153}$Eu.
Section IV determines the quadrupolar deformation of  $^{153}$Eu.
In Section V we explain the parametrisation we use for the octupolar
deformation.
Section VI describes the structure of octupole excitations.
Section VII extracts the value of octupole deformation from experimental data.
In section VIII we calculate the T- and P-odd mixing of $5/2^+$ and $5/2^-$
states in $^{153}$Eu.
EDM of $^{153}Eu$ nucleus is calculated in Section IX and SM of $^{153}Eu$
nucleus is calculated in Section X.
Section XI presents our conclusions.

\section{Experimental E1-amplitudes in $^{152}Sm$ and $^{153}Eu$}
All data in this Section are taken from Ref.~\cite{Firestone1999}.
Even-even nuclei in vicinity of  $^{153}$Eu have low energy $\approx 1$MeV
collective octupole excitation. 
There is the quadrupolar ground state rotational band  and the octupolar
rotational band starting at energy of the octupole excitation.
As a reference even-even nucleus we take $^{152}$Sm. In principle $^{154}$Sm
also would do the job, but the data for $^{154}$Sm are much less detailed,
especially on electron scattering that we discuss in Section VII.
Energies of the relevant states of the octupolar band in $^{152}$Sm 
are: $E(1^-)=963$keV, $E(3^-)=1041$keV.
The halftime of the $1^-$ state is $t_{1/2}=28.2$fs, hence the
lifetime  is $\tau=28.2/\ln(2)=40.7$fs.
The state decays via the E1-transition to the ground state, $0^+$, and to the
$2^+$ state of the ground state rotational band.
The decay branching ratio is $W(0^+)/W(2^+)=0.823$.
Therefore, the partial lifetime for $1^- \to 0^+$ transition is
  $\tau_{partial}=90$fs.
  The $1^- \to 0^+$ E1-transition decay rate is \cite{LL4}
  \begin{eqnarray}
    \label{WE1}
    \frac{1}{\tau_{partial}}=\frac{4\omega^3}{3(2j+1)}
    |\langle j^{\prime}||d||j\rangle|^2 \ ,
  \end{eqnarray}
For $1^- \to 0^+$ transition  $j=1$ and $j^{\prime}=0$. 
   The reduced matrix element of the dipole moment can be expressed in terms
  of $d_z$ in the proper reference frame of the deformed nucleus \cite{LL3}
  \begin{eqnarray}
    \label{top}
    |\langle j^{\prime}||d|| j\rangle|^2&=&\left|\sqrt{(2j+1)(2j^{\prime}+1)}
    \left(
    \begin{array}{ccc}
      j^{\prime}&1&j\\
      -m&0&m
    \end{array}\right)\right|^2\nonumber\\
&\times&|\langle 0| d_z|1\rangle|^2
  \end{eqnarray}
For $1^- \to 0^+$ transition  $j=1$,  $j^{\prime}=0$, $m=0$. Hence
 \begin{eqnarray}
 \label{Sm}  
 \langle 0| d_z|1\rangle=+ e\times 0.31fm
\end{eqnarray}
Here $e=|e|$ is the elementary charge.

$^{153}Eu$ is a deformed nucleus with the ground state $|5/2^+\rangle$.
The nearest opposite parity state $|5/2^-\rangle$ has energy $E=97.4$keV.
The halftime of the $|5/2^-\rangle$ state is $t_{1/2}=0.20$ns, hence the
lifetime is $\tau=0.29$ns. The lifetime is due to the E1-decay 
$|5/2^-\rangle \to |5/2^+\rangle$.
Using Eqs.(\ref{WE1}),(\ref{top}) with $j=j^{\prime}=m=5/2$ and comparing with
experiments we find the corresponding $d_z$ in the proper reference frame.
 \begin{eqnarray}
 \label{Eu}  
\langle 5/2^+ |d_z|5/2^-\rangle= - e\times 0.12fm
 \end{eqnarray}
Of course lifetimes do not allow to determine signs in Eqs. (\ref{Sm}) and
(\ref{Eu}). We explain in Section VI how the signs are determined.

\section{Wave functions of the ground state parity doublet $|\frac{5}{2}^{\pm})$
in $^{153}Eu$.}
The standard theoretical description of low energy states in $^{153}Eu$
is based on the Nilsson
model of a quadrupolar-deformed nucleus. In agreement with experimental data,
the model predicts the spin and parity of the ground state, $5/2^+$. It also
predicts the existence of the low-energy excited state with opposite parity,
$5/2^-$. The wave functions of the odd proton in the Nilsson scheme are
$|5/2^+\rangle=|413\frac{5}{2}\rangle$, $|5/2^-\rangle=|532\frac{5}{2}\rangle$.
Explicit form of these wave functions is presented in Appendix.
There are two rotational towers built on these states.

An alternative to Nilsson approach is the model of static collective octupolar
deformation~\cite{Flambaum2020}.
In this model the odd proton moves in the pear shape potential forming
the $\Omega=5/2$ single particle state.
A single rotational tower built on this odd proton state is consistent with
observed spectra and this is why the paper~\cite{Flambaum2020} argues in favour
of static octupole deformation. However, two different parity rotational
towers in Nilsson scheme are equally consistent with observed spectra.
Therefore, based on spectra one can conclude only that both the Nilsson model
and the static octupolar deformation model are consistent with spectra. One
needs additional data to distinguish these two models.

The Nilsson model explains the value $\Omega=5/2$ while in the
``static octupole'' model this value pops up from nowhere. However, in
principle it is possible that accidentally
the single particle state in the pear shape potential has
$\Omega=5/2$.

To resolve the issue ``Nilsson vs octupole'' we look at magnetic moments.
The magnetic moment of the ground state is $\mu_{5/2^+}=1.53\mu_N$,
see Ref.~\cite{Firestone1999}. This value is consistent with prediction
on the Nilsson model~\cite{Lamm1969}.
The magnetic moment of the $5/2^-$ state has some ambiguity,
the measurement led to two possible interpretations, 
``the recommended value'' $\mu_{5/2^-}=3.22\mu_N$, and another value
consistent with measurement $\mu_{5/2^-}=-0.52\mu_N$, see Ref.~\cite{Firestone1999}. The recommended value is consistent with the prediction of the
Nilsson model~\cite{Kemah2022}.
Thus the magnetic moments are consistent  with the Nilsson model and
inconsistent with the octupole model which implies
$\mu_{5/2^-}\approx\mu_{5/2^+}$.

While the arguments presented above rule out the static octupole model,
they do not imply that the octupole is irrelevant, actually it is relevant.
We will show now that while the Nilsson model explains magnetic moments
it cannot explain E1-amplitudes.

Within the  Nilsson model one can calculate the E1 matrix
element $\langle 5/2^+|d_z|5/2^-\rangle$.
A straightforward calculation with wave functions (\ref{Nwf}) gives the
dipole matrix element
\begin{eqnarray}
  \label{dipol}
d_z&=&e (1-Z/A)\langle 532\frac{5}{2}|z|413\frac{5}{2}\rangle\nonumber\\
  &=&e (1-Z/A)\frac{z_0}{\sqrt{2}}(0.527-0.510+0.017)\nonumber\\
  &=&e\times 0.036fm  \ .
\end{eqnarray}
Here we account the effective proton charge $(1-Z/A)=0.59$.
The calculated matrix element (\ref{dipol})
is 3 times smaller than the experimental one  (\ref{Eu}).
The first impression is that the disagreement is not bad
having in mind the dramatic compensations in Eq.(\ref{dipol}).
However, there are two following observations.\\
(i) It has been pointed out in Ref.\cite{Sushkov1984} that the compensation
in (\ref{dipol}) is not accidental: the compensation is due to the structure
of Nilsson states, and the matrix
element $ \langle 532\frac{5}{2}|z|413\frac{5}{2}\rangle$ is proportional
to the energy
splitting $\Delta E = E_{5/2^-}-E_{5/2^+}$. The matrix element is small
because $\Delta E$ is small compared to the shell model energy
$\omega_0\approx 7.7$MeV.
The value (\ref{dipol}) is calculated with wave functions from
  Ref.~\cite{BohrMottelson} that correspond to $\Delta E \approx 450$keV.
  On the other hand in reality $\Delta E \approx 97$keV.
Therefore, the true matrix element must be even smaller than the value
(\ref{dipol}).\\
(ii) The electric dipole operator is  T-even. Therefore, there is a suppression of the matrix element due to pairing of protons,  $d_z \to d_z (u_1u_2-v_1v_2)$,
where $u$ and $v$ are pairing BCS factors. This further reduces the matrix
element, see Ref.\cite{Sushkov1993}.

The arguments in the previous paragraph lead to the conclusion that while the
Nilsson model correctly predicts quantum numbers and  explains  magnetic
moments, the model does not explain the electric dipole
transition amplitude.
The experimental  amplitude is by an order of magnitude larger than the Nilsson one. This observation has been made already in Ref.\cite{Sushkov1984}.

Admixture of the collective octupole to Nilsson states resolves the dipole moment issue.
So, we take the wave functions as
\begin{eqnarray}
  \label{adm}
  |+\rangle =  |\frac{5}{2}^+\rangle &=&
  \sqrt{1-\alpha^2}|413\frac{5}{2}\rangle|0\rangle-\alpha| 532\frac{5}{2}
  \rangle|1\rangle  \nonumber\\
  |-\rangle = |\frac{5}{2}^-\rangle &=&\sqrt{1-\alpha^2}| 532\frac{5}{2}\rangle|0\rangle
  -\alpha|413\frac{5}{2}\rangle|1\rangle
\end{eqnarray}
Here the states $|0\rangle$ and $|1\rangle$ describe collective octupole mode,
$|0\rangle$ is the symmetric octupole vibration and
$|1\rangle$ is antisymmetric octupole vibration. For intuition:
$|0\rangle$ corresponds to the
ground state of $^{152}$Sm and $|1\rangle$ corresponds to the octupole
excitation at energy $\approx 1$MeV.
We will discuss in Section VI the specific structure of the states
$|0\rangle$, $|1\rangle$, explain why the mixing coefficient  in both states
in (\ref{adm}) is the same, and explain  why  $\alpha >0$.

Using (\ref{adm}) and
neglecting the small single particle contribution the transition electric dipole moment is
\begin{eqnarray}
 \label{d1}
 \langle \frac{5}{2}^+|d_z|\frac{5}{2}^-\rangle=-2\alpha \sqrt{1-\alpha^2}
 \langle 0 |d_z|1\rangle
 \end{eqnarray}
Hence, using the experimental values (\ref{Sm}) and (\ref{Eu}) we find
\begin{eqnarray}
  \label{al}
  \alpha\approx \frac{0.12}{2 \times 0.31}=0.20
  \end{eqnarray}
Thus, the weight of the admixture of the collective vibration to the simple Nilsson state is just $\alpha^2= 4\%$.
This weight is sufficiently small to make the Nilsson scheme calculation of
magnetic moments correct. On the other hand the weight is sufficiently large
to influence electric properties.

Note that the octupole vibration itself does not
have an electric  dipole transition matrix element. The E1 matrix element
is zero due to elimination of zero mode, $\langle 1|d_z|0\rangle=0$.
The nonzero value of the dipole matrix element, $\langle 1|d_z|0\rangle\ne 0$,
arises due to a small shift of the neutron
distribution with respect to the proton distribution in combination
with the octupole deformation, see e.g.
Refs.~\cite{Leander1986,Dorso1986,Butler1991}.
While this issue is important theoretically, pragmatically it is not
important to us since we take both values of matrix elements (\ref{Sm})
and (\ref{Eu}) from experiment.

It is worth noting also that in the static octupole model one expects
$\langle 5/2^+ |d_z|5/2^-\rangle= \langle 0| d_z|1\rangle=+ e\times 0.31fm$
that is like magnetic moments inconsistent with data.

\section{Quadrupolar deformation of $^{153}Eu$.}
The standard way to describe nuclear deformation is to use parameters
$\beta_l$.
In the co-rotating reference frame for the quadrupolar deformation
the surface of the nucleus is given by equation (we neglect $\beta_2^2$
compared to 1)
\begin{eqnarray}
\label{l2}
&&R(\theta)=R_0(1+\beta_2Y_{2,0})\nonumber\\
&&R_0=r_0A^{1/3}\nonumber\\
&&r_0\approx1.2fm
\end{eqnarray}
Here A is the number of nucleons.

Let us determine $\beta_2$ using the known electric quadrupole moment Q
in the ground state of $^{153}$Eu. There are two contributions in Q,
(i) collective contribution due to the collective deformation,
(ii) single particle contribution of the odd proton. Using Nilsson
wave functions it is easy to check that the single particle contribution is
about 3-4\% of the experimental one, so it can be neglected.
Collective electric quadrupole moment is given by density of protons $\rho_p$,
\begin{eqnarray}
  \label{qm}
  Q_0=Q_{zz}&=&\int\rho_p(3z^2-r^2)dV=4\sqrt{\frac{\pi}{5}}\int\rho_pr^2Y_{20}dV
  \nonumber\\
  &=&\frac{3ZR_0^2}{\sqrt{5\pi}}\beta_2
  \left[1+\frac{2\sqrt{5}}{7\sqrt{\pi}}\beta_2
    +\frac{12}{7\sqrt{\pi}}\beta_4\right]
    \end{eqnarray}
Here we also account $\beta_4$. Z is the nuclear charge.
Eq.(\ref{qm}) gives the quadrupole moment in the proper reference frame.
In the laboratory frame for the ground state, $J=\Omega=5/2$, the quadrupole
moment is $Q=\frac{5}{14}Q_0$, see problem to $\S 119$ in Ref.\cite{LL3}.
The ground state quadrupole moment of $^{153}$Eu is
$Q=2.412$ barn~\cite{Firestone1999}. From here, assuming $\beta_4=0.07$,
we find the quadrupole deformation of $^{153}$Eu nucleus in the ground state,
\begin{eqnarray}
 \label{b2}
\beta_2\approx 0.29 \ .
\end{eqnarray}
The values $\beta_2\approx 0.29$, $\beta_4=0.07$ perfectly agree with that
in $^{152}$Sm determined from electron scattering~\cite{Bertozzi1972}.

The electric quadrupole moment of $^{151}$Eu nucleus in the ground state is
$Q=0.903$barn~\cite{Firestone1999}.
Therefore, in $^{151}$Eu the quadrupolar deformation, $\beta_2\approx 0.12$,
is significantly smaller than that in  $^{153}$Eu.

\section{Nuclear density variation due to  the octupole deformation}
The standard way to describe the static octupole deformation $\beta_3$ is to
use parametrisation (\ref{l2})
\begin{eqnarray}
\label{l3}
R(\theta)=R_0(1+\beta_1Y_{10}+\beta_2Y_{2,0}+\beta_3Y_{3,0}+...)
\end{eqnarray}
This Eq. describes the surface of nucleus in the proper reference frame.
The dipole harmonic $Y_{10}$ is necessary to eliminate the zero mode, i.e.
to satisfy the condition
\begin{eqnarray}
  \label{zerom}
  \langle z\rangle= \int\rho(r)rY_{10}dV=0
  \end{eqnarray}
  where $\rho(r)$ is the number density of nucleons.
From (\ref{zerom}) we find 
\begin{eqnarray}
  \label{b1}
\beta_1=-x\beta_2\beta_3 \ , \ \ \ x=\sqrt{\frac{243}{140\pi}}\approx 0.743 \ .
\end{eqnarray}
For our purposes it is more convenient to use parametrisation different from
(\ref{l3}), the parametrisation we use is
\begin{eqnarray}
  \label{l33}
  \delta\rho= \beta_3
  \frac{3A}{4\pi R_0^2}\delta[r-R_0(1+\beta_2Y_{20})](Y_{30}-x\beta_2Y_{10}) \ .
\end{eqnarray}
Here $\delta \rho$ is the octupolar component of the nuclear density.
Due to the $\delta$-function, $\delta[...]$,  the component is nonzero only
at the surface  of the nucleus. Parametrisations (\ref{l3}) and (\ref{l33})
are equivalent, both satisfy the constraint (\ref{zerom}) and
both give the same octupole moment
\begin{eqnarray}
  \label{oct}
  Q_{30}=\sqrt{\frac{4\pi}{7}}\int\rho r^3 Y_{30}dV=
  \beta_3\frac{3A}{\sqrt{28\pi}}R_0^3 \ .
 \end{eqnarray}   

\section{Structure of the vibrational states $|0\rangle$, $|1\rangle$}
The deformation picture described in the previous section is purely classical.
Quantum mechanics makes some difference.
We work in the proper reference frame where the nuclear axis direction, the
z-axis, is fixed.
%%%%%%%%%%%%%%%%%
\begin{figure}[h!]
\includegraphics[width=0.3\textwidth]{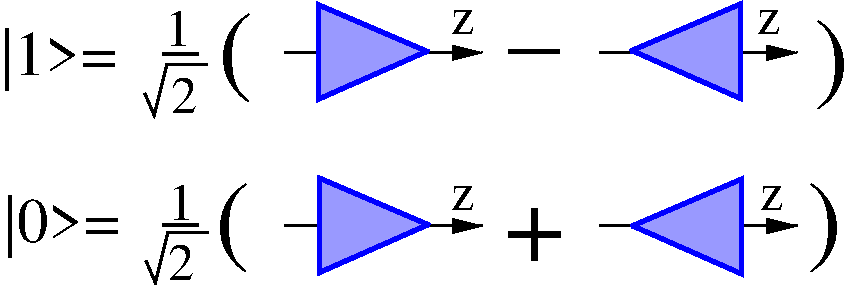}
\caption{The states $|0\rangle$ and  $|1\rangle$.
}
\label{deform}
\end{figure}
%%%%%%%%%%%%%%%%%
Hence, there are two possible orientations of the pear, as it is shown in
Fig.\ref{deform}. There is tunnelling between these two orientations, the
tunnelling leads to the energy splitting and to formations of
symmetric and antisymmetric states  
$|0\rangle$, $|1\rangle$. This picture is valid when the tunnelling energy
splitting, $\Delta E_{tun}$,  is larger than the rotational energy
splitting,
$\Delta E_{rot}$. Experimentally $\Delta E_{tun}\sim 1$MeV,
$\Delta E_{rot}\approx 20$keV, so the description is well justified.
The description of Fig.~\ref{deform} implies that
  the octupole  deformation is quasistatic. The quasistatic description is
  justified by the existence of well defined rotational towers in $^{152}$Sm
  built on $|0\rangle$ and $|1\rangle$ states, see Ref.~\cite{Firestone1999}.
  Note that even if the pear tunneling amplitude is comparable with the
  rotatinal energy, $\Delta E_{tun}\sim \Delta E_{rot}$, the octupole
  deformation  is not static. To have a trully static octupole one needs 
 $\Delta E_{tun}\ll \Delta E_{rot}$.

The Hamiltonian for the odd proton reads
\begin{eqnarray}
  \label{Hsc}
  H=\frac{p^2}{2m}+U(r) \ .
\end{eqnarray}
Here $U(r)$ is the selfconsistent potential of the even-even core.
It is well known that the nuclear density $\rho(r)$ has approximately the
same shape  as the potential
\begin{eqnarray}
  \label{shape}
U(r) \approx \frac{U_0}{\rho(0)}\rho(r) \ ,  
\end{eqnarray}
where $U_0\approx -50$MeV and $\rho(0)=3/(4\pi r_0^3)$.
  Hence the variation of the potential related to the octupole deformation is
  \begin{eqnarray}
    \label{dU}
    \delta U &=& \frac{U_0}{\rho(0)}\delta \rho\\
    &=&    
    \beta_3 U_0 R_0 \delta[r-R_0(1+\beta_2Y_{20})](Y_{30}-x\beta_2Y_{10})\ .
    \nonumber
  \end{eqnarray}
  This  is the perturbation that mixes single particle Nilssen states
  with simultaneous mixing of $|0\rangle$ and $|1\rangle$. The mixing matrix
  element is 
  \begin{eqnarray}
   \label{mix}
   &&M=\langle 1|\langle 532\frac{5}{2}|\delta U|413\frac{5}{2}\rangle|0\rangle
   = \int \rho_{sp}(r)\delta U(r) dV\nonumber\\
&&  \rho_{sp}(r)= \langle\psi^*_{532}(r)\psi_{413}(r)\rangle \ .  
 \end{eqnarray}
 Here $\rho_{sp}$ is offdiagonal single particle density
 of Nilsson wave functions  (\ref{Nwf}),
the density depends on $r$, the brackets $\langle ..\rangle$ in
$\rho_{sp}$  denote averaging over spin only.
  Numerical evaluation of the mixing matrix element is straightforward,
 the answer at $\beta_2=0.29$ is $M\approx 5\beta_3$MeV. The value slightly
 depends on $\beta_2$, at $\beta_2=0$ the value of M is 10\% smaller.
  The coefficient $\alpha$ in Eqs.(\ref{adm}) is
 \begin{eqnarray}
   \label{al1}
   \alpha=\frac{M}{\Delta E_{tun}} \ ,
 \end{eqnarray}
 where $\Delta E_{tun}\approx 1$MeV.
 Eqs.(\ref{al1}),(\ref{mix}) together with positive value of M explain why the
 coefficient $\alpha$ is the same in both   Eqs.(\ref{adm}) and why
 $\alpha > 0$.

 Moreover, comparing (\ref{al1}) with value of $\alpha$ extracted from
 experimental data, Eq.(\ref{al}), we determine the octupole deformation,
 $\beta_3 =0.04$. While the value is reasonable, unfortunately one cannot
 say that this is the accurate value.
 The shape approximation (\ref{shape}) is not very accurate.
Even more important, it is not clear how the BCS factor influences $\rho_{sp}$. The BCS factor can easily
 reduce $\rho_{sp}$  by factor $\sim 2-3$, hence increasing $\beta_3$ by the
 same factor.
Theoretical calculations of $\beta_3$ give
values from 0.05~\cite{Butler1991}, to 0.075~\cite{Ebata2017}, and
even 0.15~\cite{Zhang2010}.

\section{The value of the octupole deformation parameter $\beta_3$}

With wave functions shown in Fig.\ref{deform} one immediately finds
the electric octupole  matrix element between states $|0\rangle$ and
$|1\rangle$
\begin{eqnarray}
  \label{QE3}
\langle 1|Q_{30}^{(e)}|0\rangle=e\frac{Z}{A}Q_{30} \ ,
\end{eqnarray}
where $Q_{30}$ is given by Eq.(\ref{oct}).
We are not aware of direct  measurements of $Q_{30}^{(e)}$ in $^{152}Sm$.
The book~\cite{BohrMottelson} presents the ``oscillator strengths'' for
corresponding E3 transitions in $^{152}$Sm and $^{238}$U,
$^{152}Sm: \ B_3=1.2\times 10^5e^2fm^6$, $^{238}U: \ B_3=5\times 10^5e^2fm^6$
(table 6.14 in the book).
However, these values have been determined not from direct electromagnetic
measurements, the ``oscillator strengths'' have been indirectly  extracted from
deuteron scattering from the nuclei.
Fortunately for $^{238}$U there is a more recent value determined from the
electron scattering~\cite{Hirsch1978}: $B_3=(6.4\pm 0.6)\times 10^5e^2fm^6$.
All in all this data give $\beta_3 \approx 0.08$ for both $^{152}$Sm and
$^{238}$U.

\begin{figure}[h!]
\includegraphics[width=0.5\textwidth]{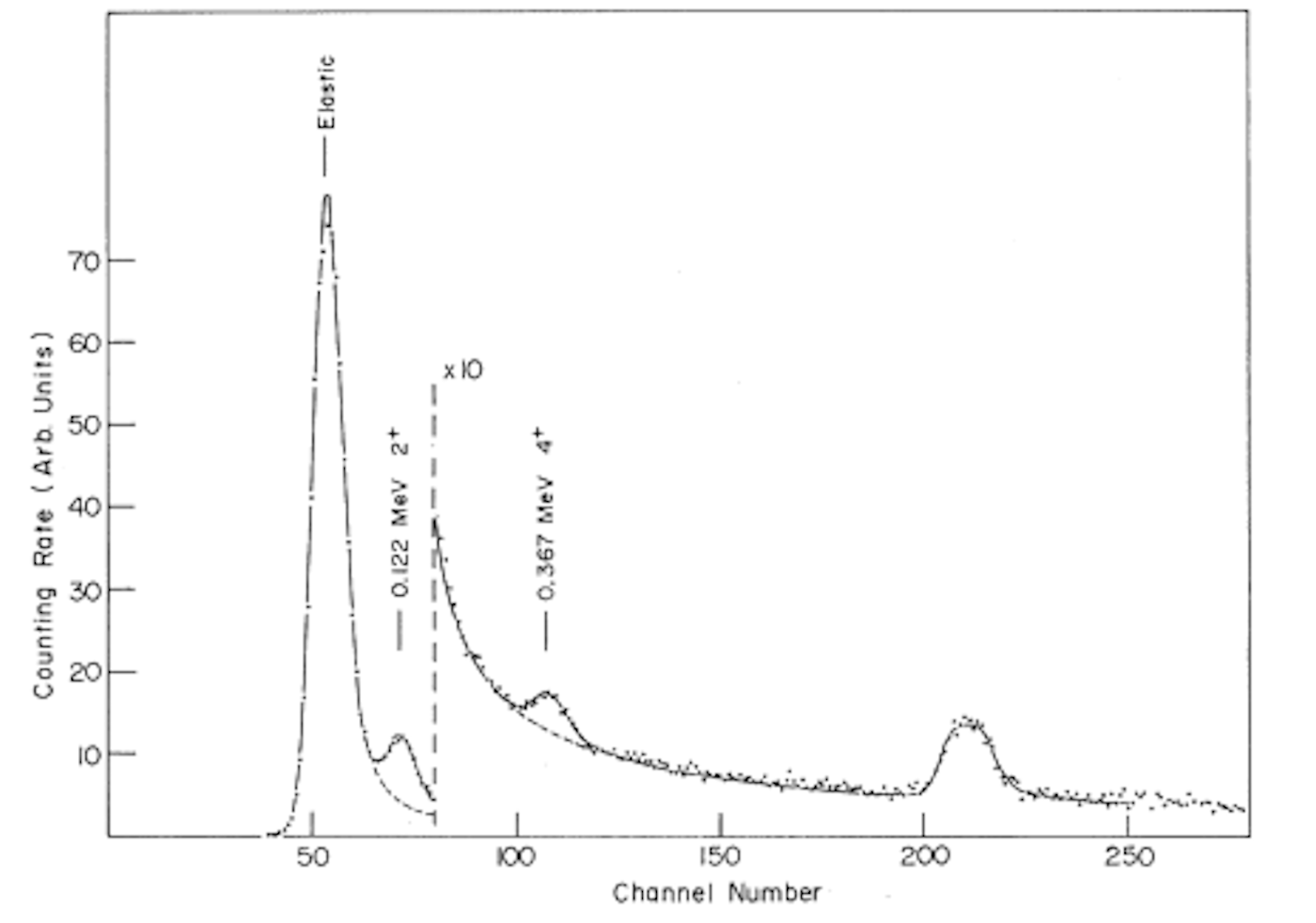}
\caption{Spectrum of scattered electrons from $^{152}$Sm at $93.5^o$.
Incident electron energy, 76 MeV. Besides the
ground-state rotational band, the $3^-$ level at 1. 041 MeV and the $2^+$
level at 1.086 MeV are seen (channel 210). The plot is taken from
Ref.~\cite{Bertozzi1972}
}

\label{Bert}
\end{figure}

\begin{figure}[h!]
\includegraphics[width=0.23\textwidth]{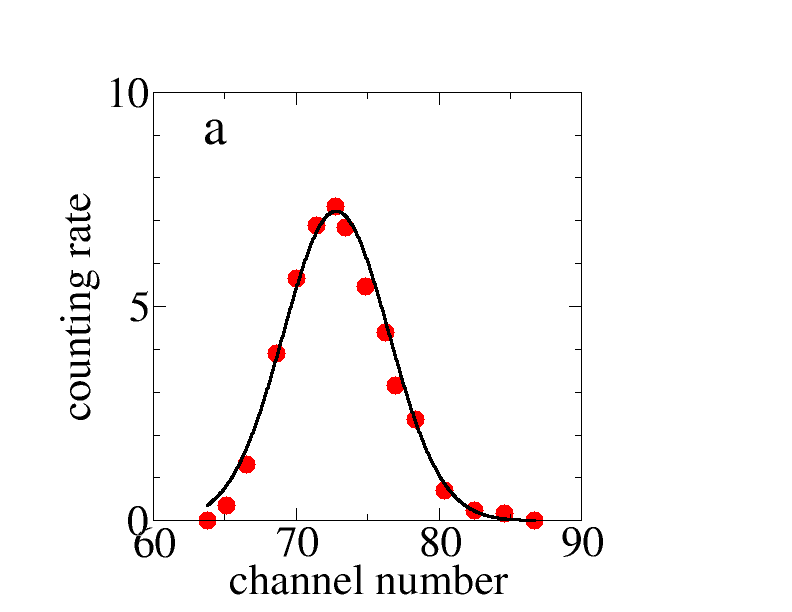}
\includegraphics[width=0.23\textwidth]{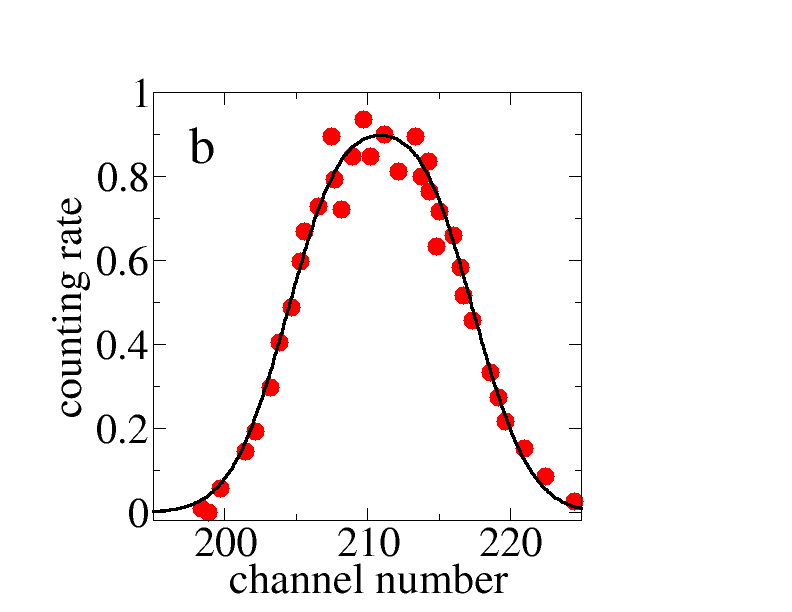}
\caption{Excitation peaks with subtracted background. Red dots are
  experimental data and black curves are Gaussian (double Gaussian)  fits.
  Panel a: The $2^+$ excitation of  the ground state rotational band.
  Panel b: The combined peak of the $3^-$ octupole and the $\gamma 2^+$ state of
  the rotational $\gamma$-band.
  }
\label{Bert1}
\end{figure}

 Fortunately, the electron scattering data~\cite{Bertozzi1972}
allow to determine $\beta_3$ in $^{152}$Sm pretty accurately.
The Ref.~\cite{Bertozzi1972} was aimed to determine $\beta_2$ and $\beta_4$,
their results, $\beta_2=0.287\pm 0.003$, $\beta_4=0.070\pm 0.003$ are remarkably
close to that we get for $^{153}$Eu in Section IV.
The inelastic scattering spectrum copied from Ref.~\cite{Bertozzi1972} is
shown in Fig.\ref{Bert}.
%%%%%%%%%%%%%%%%%
%%%%%%%%%%%%%%%%%
Here we reanalyse the spectrum.
The first inelastic peak at E=122keV ($\approx$ channel 73) corresponds to
the  $2^+$
excitation of the rotational ground state band. The peak after subtraction
of the background is shown in panel a of Fig.\ref{Bert1}.
%%%%%%%%%%%%%%%%%
%%%%%%%%%%%%%%%%%
Red dots are experimental points and the solid curve is the Gaussian fit
\begin{eqnarray}
  \label{fG1}
&& I=Ae^{-(x-x_0)^2/\sigma^2}\nonumber\\
  &&A=7.23,  \ x_0=72.9, \  \sigma=5.21 \ .
\end{eqnarray}
Hence, the halfwidth is
$\Gamma =2\ln(2)\sigma=49.3$keV. Here we account that one channel step is
6.82keV.
This energy resolution is 0.065\% of electron energy 76MeV.
This is slightly smaller than the ``typical value'' 0.08\% mentioned
in Ref.~\cite{Bertozzi1972}.
The peak at Fig.\ref{Bert} near the channel 210 is a combination of the
$3^-$ octupole (E=1041keV) and  of the $\gamma 2^+$ state of the
$\gamma$-band (E=1086keV).
The peak after subtraction of the background is shown in panel b of
Fig.\ref{Bert1}. We fit the double peak by the  double Gaussian 
\begin{eqnarray}
  \label{fG2}
  &&  I=B\left[e^{-(x-x_1)^2/\sigma^2}+e^{-(x-x_2)^2/\sigma^2}\right]\nonumber\\
  &&B=0.670,  \ x_1=207.6, \ x_2=214.2, \ \sigma=5.21.
\end{eqnarray}
The value of $x_1$ corresponds to E=1041keV, the value of $x_2$ corresponds
to E=1086keV, $\sigma$ is known from (\ref{fG1}).
The fitting shows that intensities of $3^-$ and  $\gamma 2^+$ lines cannot
differ by more than 5\%, so we take them equal. Therefore in the end
there is only one fitting parameter B.

Based on Eqs.(\ref{fG1}) and (\ref{fG2}) we find the ratio of spectral weights
\begin{eqnarray}
  \label{rsp}
  \frac{S(3^-)}{S(2^+)}=\frac{B}{A}=0.093
\end{eqnarray}
Here $2^+$ is the ground state rotational state.
Interestingly, the analysis gives also the spectral weight of the
$\gamma 2^+$ state. This allows to determine the magnituide of the
$\gamma$-deformation.  However, this issue is irrelevant to the Schiff
moment and therefore we do not analyse it further.

Coulomb potential or Eu nucleus at $r \approx R_0$ is 15MeV. This is
significantly smaller than the electron energy 76MeV. Therefore, the electron
wavefunction can be considered as a plane wave. The momentum transfer is
\begin{eqnarray}
  \label{qt}
  q=2p\sin(93.5^o/2)\approx 111MeV\approx 0.562fm^{-1} \ .
\end{eqnarray}
Using expansion of the plane wave in spherical harmonics together with
Wigner-Eckart theorem the spectral weights can be expressed as integrals
in the co-rotating reference frame
\begin{eqnarray}
  \label{s23}
  S(2^+) &\propto& \left|\int Y_{20}j_2(qr)\rho(r) dV\right|^2\nonumber\\
  S(3^-) &\propto& \left|\int Y_{30}j_3(qr)\delta\rho(r) dV\right|^2
  \end{eqnarray}
Here $j_l(qr)$ is the spherical Bessel function~\cite{LL3},
$\rho(r)$ is the density with quadrupolar deformation, and
$\delta\rho$ is given by Eq.(\ref{l33}).
The coefficient of proportionality in both equations (\ref{s23}) is the same
and therefore we skip it. Evaluation of integrals in (\ref{s23})
is straightforward, it gives
\begin{eqnarray}
  \label{s23a}
  &&\int Y_{20}j_2(qr)\rho(r) dV \propto \beta_2j_2(qR_0)=0.302\beta_2\nonumber\\
&&\int Y_{30}j_3(qr)\delta\rho(r) dV \propto \beta_3j_3(qR_0)=0.205\beta_3
  \end{eqnarray}
Comparing the theoretical ratio with it's experimental value (\ref{rsp}) and
using the known quadrupolar deformation we find the octupolar deformation
$\beta_3=0.45\beta_2=0.130$.

In the previous paragraph we used the plane wave approximation
for the electron wave function neglecting the Coulomb potential $\approx 15$MeV
compared to the electron energy 76MeV. A simple way to estimate the
Coulomb correction is to change $q\to q^{\prime}\approx q(1+15/76)=0.673fm^{-1}$.
This results in  $\beta_3=0.090$. Probably this simplistic way overestimates the
effect of the Coulomb potential. An accurate calculation of distorted electron
wave functions would allow to determine $\beta_3$ very accurately.
For now we take
\begin{eqnarray}
  \label{b3}
  \beta_3= 0.10
\end{eqnarray}

\section{T- and P-odd mixing of $5/2^+$ and $5/2^-$ states in $^{153}Eu$}
The operator of the T, P-odd interaction reads~\cite{Sushkov1984}
\begin{eqnarray}
  \label{TP}
  H_{TP}=\eta \frac{G}{2\sqrt{2}m} {\vec \sigma}\cdot{\vec \nabla}\rho
\end{eqnarray}
Here $G\approx 1.03/m^2$ is the Fermi constant, $\eta$ is a dimensionless
constant characterising the interaction, ${\vec \sigma}$ is the Pauli matrix
corresponding to the spin of unpaired nucleon, and $\rho$ is the nuclear
number density.
The single particle matrix element of $H_{TP}$ between the Nilsson states can be
estimated as, see Ref.~\cite{Sushkov1984},
\begin{eqnarray}
 && \langle 532|H_{TP}|413\rangle
\propto 
\langle 532|\nabla \rho|413\rangle
\propto 
\langle 532|\nabla U|413\rangle\nonumber\\
&&\propto
\langle 532|[p, H]|413\rangle
\propto 
(E_{532}-E_{413}) \langle 532|p|413\rangle\nonumber\\
&&\propto 
(E_{532}-E_{413}) \langle 532|[r,H]|413\rangle\nonumber\\
&&\propto 
(E_{532}-E_{412})^2 \langle 532|r|413\rangle
\nonumber
  \end{eqnarray}
Thus, the  matrix element is suppressed by the small parameter $(\Delta E/\omega_0)^2$,
with $\Delta E \approx 100$keV and $\omega_0 \approx 8$MeV. Hence,
the single particle matrix element can be neglected.

\begin{figure}[h!]
\includegraphics[width=0.25\textwidth]{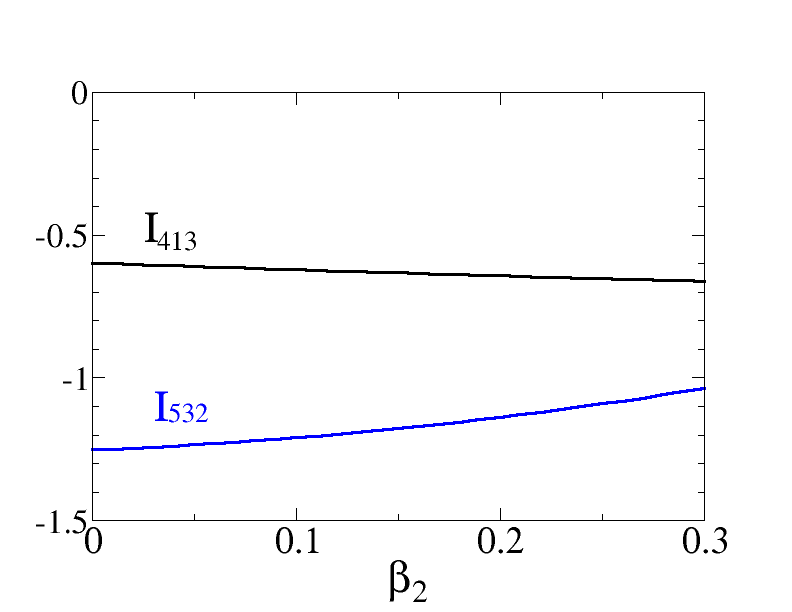}
\caption{Dimensionless matrix elements $I_{413}$ and $I_{532}$ vs quadrupolar deformation. 
}
\label{II}
\end{figure}

The matrix element between the physical states (\ref{adm}) contains also the
collective octupole contribution
\begin{eqnarray}
  \label{TP1}
\langle - |H_{TP}|+\rangle=&&  -\alpha\langle 532\frac{5}{2}|\langle 1|H_{TP}|
0\rangle|532\frac{5}{2}\rangle \nonumber\\   
&&  -\alpha\langle 413\frac{5}{2}|\langle 0|H_{TP}|
1\rangle|413\frac{5}{2}\rangle    
\end{eqnarray}
Integrating by parts we transform this to
\begin{eqnarray}
  \label{TP2}
  \langle - |H_{TP}|+\rangle&=&  \frac{\alpha\eta G}{2\sqrt{2}m}
  \int \left[\rho_{532}(r)+\rho_{413}\right]\delta\rho (r)  dV\nonumber\\
\rho_{532}(r)&=& \partial_z\langle 532|\sigma_z|532\rangle\nonumber\\
\rho_{413}(r)&=& \partial_z\langle 413|\sigma_z|413\rangle
\end{eqnarray}
Here $\delta\rho$ is the octupole density (\ref{l33}). Note that the ``spin densities''
$\rho_{532}$ and $\rho_{413}$  depend on $r$, the brackets $\langle ..\rangle$ in
definition of the densities in (\ref{TP2}) denote averaging over spin only.
Note also that the ``spin densities'' are T-odd. Therefore, the BCS factor
practically does not influence them.
Numerical evaluation of integrals in (\ref{TP2}) with Nilsson wave functions (\ref{Nwf}) is straightforward, the result is
\begin{eqnarray}
  \label{TP3}
  \langle - |H_{TP}|+\rangle&=& \alpha\eta\beta_3
  \frac{ G}{2\sqrt{2}m}\frac{3A}{4\pi R_0^4}[ I_{413}+I_{532}]
\end{eqnarray}
Dimensionless $I_{413}$ and $I_{532}$ are plotted in Fig.\ref{II} versus $\beta_2$.
%%%%%%%%%%%%%%%%%
%%%%%%%%%%%%%%%%%
At the physical deformation, $\beta_2=0.29$, Eq.(\ref{b2}), the values are
$I_{413}=-0.66$ and $I_{532}=-1.05$. Hence we arrive to the following mixing
matrix element
\begin{eqnarray}
  \label{TP4}
  \langle - |H_{TP}|+\rangle=-0.24  \alpha  \eta \beta_3\ eV \ .
\end{eqnarray}
%This equation, besides the coefficient $\alpha$,  agrees in magnitude
%with the estimate given in Eq.(7) of Ref.~\cite{Flambaum2020},
%$  \langle - |H_{TP}|+\rangle=0.18  \eta \beta_3\ eV$.
%The sign is opposite.

\section{Electric dipole moment of $^{153}Eu$ nucleus}
\begin{figure}[h!]
\includegraphics[width=0.15\textwidth]{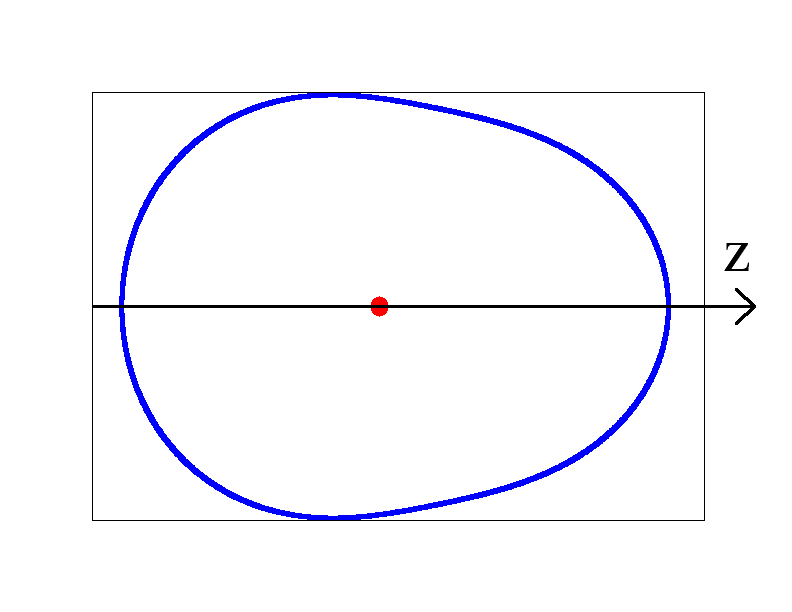}
\caption{Pear shape with $\beta_2=0.29$, $\beta_3=0.10$
}
\label{pear}
\end{figure}

We need to determine signs in Eq.(\ref{Sm}) and Eq.(\ref{Eu}).
In our notations $\beta_3 > 0 $  corresponds to the pear orientation with
respect to the z-axis shown in Fig.\ref{pear}.
%%%%%%%%%%%%%%%%%
  According to Refs.~\cite{Leander1986,Dorso1986,Butler1991} protons are
  shifted in the positive z-direction. Hence, $d_z$ in Eq.(\ref{Sm}) is
  positive. Hence, using Eqs.(\ref{adm}), we conclude that the sign in
  Eq.(\ref{Eu}) is negative.

With Eqs.(\ref{TP4}) and (\ref{Eu}) we find the T,P-odd electric dipole moment
in the ground state.
\begin{eqnarray}
  \label{dTP}
  d^{TP}_z&=&2\frac{\langle+|d_z|-\rangle\langle-|H_{TP}|+\rangle}{E_+-E_-}
  \nonumber\\
&=&-0.59\times 10^{-6}\alpha\beta_3\eta [e \cdot fm]\nonumber\\
  &=& -1.18\times 10^{-8}\eta [e\cdot fm] \ .
\end{eqnarray}
For the numerical value we take $\alpha=0.20$, see Eq.(\ref{al}), and
$\beta_3=0.10$, see Eq.(\ref{b3}).
Eq. (\ref{dTP}) gives  the EDM in the co-rotating reference frame. The EDM in
the laboratory reference frame is
\begin{eqnarray}
  \label{dTP1}
  d^{TP}=\frac{5}{7}d^{TP}_z = -0.84\times 10^{-8}\eta [e \cdot fm]
\end{eqnarray}
This EDM is comparable with that of a heavy spherical nucleus, see
Ref.~\cite{Sushkov1984}

\section{Schiff moment of $^{153}Eu$ nucleus}
The operator of the Schiff moment (SM) reads~\cite{Sushkov1984}
\begin{eqnarray}
  \label{sch1}
  {\hat S}_z=\frac{1}{10}\left[\int \rho_q r^2 z dV -\frac{5}{3}r_q^2d_z\right]
\end{eqnarray}
It is a vector. Here $\rho_q$ is the charge density and
\begin{eqnarray}
r_q^2 \approx \frac{3}{5}R_0^2  
\end{eqnarray}
is the rms electric charge radius squared.
With the  static octupole deformation (\ref{l33})  the 1st
term in (\ref{sch1}) is 
\begin{eqnarray}
  \label{sch2}
S_{intr}=\frac{1}{10}\int \rho_q r^2 z dV =\frac{9}{20\sqrt{35}\pi}eZR_0^3\beta_2\beta_3
\end{eqnarray}
Here we use the same notation $S_{intr}$ as that in
Refs.\cite{Spevak1997,Flambaum2020}.
%Eq.(\ref{sch2})  agrees  Ref.\cite{Spevak1997}.
%but it is 3 times larger than Eq.(3) from Ref.~\cite{Flambaum2020}
%It seems that the factor 3 is missing in Ref.~\cite{Flambaum2020}
The matrix element of the first term in (\ref{sch1}) between the states
(\ref{adm}) is
\begin{eqnarray}
\label{sch3}
\langle +|{\hat S}_{1z}|-\rangle& =&-2\alpha S_{intr}\nonumber\\
&=&-\alpha\frac{9}{10\pi\sqrt{35}}eZR_0^3\beta_2\beta_3
\end{eqnarray}
Combining this with Eq.(\ref{TP4}) we find the expectation value over the
ground state
\begin{eqnarray}
\label{sch4}
\langle +|{\hat S}_{1z}|+\rangle &=&
2\frac{\langle+|{\hat S}_{1z}|-\rangle\langle-|H_{TP}|+\rangle}{E_+-E_-}\nonumber\\
&=&-0.24\times 10^{-6}e Z R_0^3\alpha^2\beta_2\beta_3^2\eta
\end{eqnarray}
Hence, the Schiff moment is
\begin{eqnarray}
\label{sch5}
S_z &=&\langle +|{\hat S}_z|+\rangle =\langle +|{\hat S}_{1z}|+\rangle-
\frac{1}{10}R_0^2d_z^{TP}
\nonumber\\
&=&\left[-4.0\times 10^{-3}\alpha^2\beta_2\beta_3^2
  +2.4\times 10^{-6} \alpha\beta_3\right]\eta [e \cdot fm^3]\nonumber\\
&=&-4.16\times10^{-7}\eta [e \cdot fm^3]
\end{eqnarray}
For the final numerical value we take $\alpha=0.20$, see Eq.(\ref{al}),
$\beta_2=0.29$, see Eq.(\ref{b2}) and $\beta_3=0.10$, see Eq.(\ref{b3}).
Note that the first term in the middle line of Eq.(\ref{sch5}) is proportional
to $\alpha^2\beta_3^2$ and at the same time the second term is 
proportional to $\alpha\beta_3$. This is because one power of $\alpha\beta_3$
is ``hidden'' in the experimental dipole matrix element (\ref{Eu}).
The second term is just about 10\% of the first one.
Eq. (\ref{sch5}) gives  the Schiff moment  in the co-rotating reference frame.
The Schiff moment in the laboratory reference frame is
\begin{eqnarray}
  \label{sch6}
  S=\frac{5}{7}S_z = -2.97\times 10^{-7}\eta [e \cdot fm^3]
\end{eqnarray}
This result is pretty reliable, the major uncertainty about
factor 2 is due to uncertainty in the value of $\beta_3$.
A more accurate analysis of inelastic electron scattering
data~\cite{Bertozzi1972}, see Section V,  can reduce the uncertainty.

In $^{151}$Eu the energy splitting $E_--E_+$ is 3.5 times larger than that
in $^{153}$Eu, and the quadrupolar deformation is 2.5 times smaller.
Therefore, the Schiff moment is at least an order of magnitude smaller
than that of $^{153}$Eu.
Unfortunately, there is no enough data for an accurate calculation for
$^{151}$Eu.

Another interesting deformed nucleus is $^{237}$Np.
  Performing a simple rescaling from our result for $^{153}$Eu we get the
  following estimate  of $^{237}$Np  Schiff Moment,
  $S \sim-1.5\times 10^{-6}\eta [e \cdot fm^3]$. This is 40 times larger than
  the   single particle estimate~\cite{Sushkov1984}.
  Of course following our method  and using $^{238}$U as a  reference nucleus
(like the pair $^{153}$Eu, $^{152}$Sm in the present work) one can
  perform an accurate calculations of $^{237}$Np  Schiff moment.  
  Data for $^{238}$U are available in Ref.~\cite{Hirsch1978}.

\section{Conclusions}
The Hamiltonian of nuclear time and parity violating interaction is
defined by Eq.(\ref{TP}). For connection of the dimensionless
interaction constant $\eta$ with the QCD axion $\theta$-parameter see
Ref.~\cite{ASushkov2023}. The Hamiltonian (\ref{TP}) leads to the Schiff moment
of a nucleus.
In the present work we have dveloped a new method to calculate Schiff moment
of an even-odd deformed nucleus.
 The method is essentially based on experimental data  on magnetic moments and
  E1,E3-amplitudes in the given even-odd nucleus and in adjacent even-even
  nuclei. Unfortunately such sets of data are not known yet for most of
  interesting nuclei.
Fortunately the full set of necessary data exists for $^{153}$Eu.
  Hence, using the new method, we perform the calculation for $^{153}$Eu.
   The result is given by Eq.(\ref{sch6}).
The theoretical uncertainty of this result, about factor 2, is mainly due to
the uncertainty in the value of the octupole deformation.
A more sophisticated  analysis of available electron scattering
data  can further reduce the uncertainty.

The Schiff Moment (\ref{sch6}) is about 20-50 times larger than that in
heavy spherical nuclei~\cite{Sushkov1984} and it is
3 times larger than what paper~\cite{ASushkov2023}
calls ``conservative estimate''.
On the other hand it is  by factor 30 smaller than the result of
Ref.~\cite{Flambaum2020} based on the model of static octupole deformation.

Using the calculated value of the Schiff Moment we  rescale results of
Ref.~\cite{ASushkov2023} for the energy shift of $^{153}$Eu nuclear spin
and for the effective electric field in EuCl$_3\cdot$ 6H$_2$O compound.
The result of the rescaling is
\begin{eqnarray}
  &&\delta {\cal E}_o= 0.9\times 10^{-9}\theta [eV]\nonumber\\
  &&E_o^*=0.3 MV/cm
\end{eqnarray}
These are figures of merit for the proposed~\cite{ASushkov2023a}
 Cosmic Axion Spin Precession Experiment with EuCl$_3\cdot$ 6H$_2$O.

\section*{Acknowledgement}
I am grateful to A. O. Sushkov for stimulating discussions and interest to the
work. This work has been supported by the Australian Research Council Centre
of Excellence in Future Low-Energy Electronics Technology (FLEET)
(CE170100039).

\appendix

\section{Nilsson wave functions}
Parameters of the deformed oscillator potential used in  Nilsson model are
\begin{eqnarray}
  \label{osc}
  &&  \omega_z=\omega_0\left(1-\frac{2}{3}\delta\right) , \ \ \ \
  z_0=\frac{1}{\sqrt{m\omega_z}}\nonumber\\
  &&  \omega_{\rho}=\omega_0\left(1+\frac{1}{3}\delta\right) , \ \ \ \
  \rho_0=\frac{1}{\sqrt{m\omega_{\rho}}}\nonumber\\
  &&\omega_0=\frac{41MeV}{A^{1/3}}
\end{eqnarray}
where $m\approx 940$MeV is the nucleon mass.
The parameter  $\delta$ is related $\beta_2$  used in the main text
as
\begin{eqnarray}
  \delta=\frac{3\sqrt{5}}{4\sqrt{\pi}}\beta_2\approx 0.946\beta_2 \ .
  \end{eqnarray}
The oscillator wave functions defined in Ref.~\cite{BohrMottelson} are
\begin{eqnarray}
  \label{oscwf}
  {\overline z}&=&z/z_0\nonumber\\
  |0\rangle_z&=&\frac{1}{(\sqrt{\pi}z_0)^{1/2}}e^{-{\overline z}^2/2}\nonumber\\
  |1\rangle_z&=&\frac{\sqrt{2}}{(\sqrt{\pi}z_0)^{1/2}}{\overline z}
  e^{-{\overline z}^2/2}\nonumber\\
  |2\rangle_z&=&\frac{1}{(2\sqrt{\pi}z_0)^{1/2}}[2{\overline z}^2-1]
  e^{-{\overline z}^2/2}\nonumber\\
  |3\rangle_z&=&\frac{1}{(3\sqrt{\pi}z_0)^{1/2}}{\overline z}[2{\overline z}^2-3]
  e^{-{\overline z}^2/2}\nonumber\\
  {\overline \rho}&=&\rho/\rho_0\nonumber\\
  |2,2\rangle_{\rho}&=&
  \frac{1}{\sqrt{2\pi}\rho_0}{\overline \rho}^2e^{-{\overline \rho}^2/2}e^{2i\varphi}
  \nonumber\\
  |3,3\rangle_{\rho}&=&
  \frac{1}{\sqrt{6\pi}\rho_0}{\overline \rho}^3 e^{-{\overline \rho}^2/2}e^{3i\varphi}\nonumber\\
  |4,2\rangle_{\rho}&=&
  \frac{1}{\sqrt{6\pi}\rho_0}{\overline \rho}^2({\overline \rho}^2-3)
  e^{-{\overline \rho}^2/2}e^{2i\varphi}\nonumber\\
  |5,3\rangle_{\rho}&=&
  \frac{1}{\sqrt{24\pi}\rho_0}{\overline \rho}^3({\overline \rho}^2-4)
  e^{-{\overline \rho}^2/2}e^{3i\varphi}
\end{eqnarray}
The Nilsson wave functions for the quadrupolar deformation $\delta=0.3$
written the oscillator basis (\ref{oscwf}) are \cite{BohrMottelson}
\begin{eqnarray}
  \label{Nwf}
|413\frac{5}{2}\rangle&=&0.938|1\rangle_z|3,3\rangle_{\rho}|\downarrow\rangle
  -0.342|2\rangle_z|2,2\rangle_{\rho}|\uparrow\rangle\nonumber\\
  &+&0.054|0\rangle_z|4,2\rangle_{\rho}|\uparrow\rangle\\
|532\frac{5}{2}\rangle&=&
  0.861|3\rangle_z|2,2\rangle_{\rho}|\uparrow\rangle
  +0.397|2\rangle_z|3,3\rangle_{\rho}|\downarrow\rangle\nonumber\\
  &+&0.310|1\rangle_z|4,2\rangle_{\rho}|\uparrow\rangle
+0.075|0\rangle_z|5,3\rangle_{\rho}|\downarrow\rangle\nonumber
\end{eqnarray}

\end{document}